\documentclass[preprint,authoryear,12pt]{elsarticle}

\usepackage{epsfig}

\usepackage{amssymb}

\usepackage[%ps2pdf,%
a4paper=true,%
breaklinks=true,%
colorlinks=true,%
pdfauthor={First Author et al.},%
pdftitle={Template for manuscripts in Advances in Space Research}%
]{hyperref}

\usepackage{graphicx}
\usepackage{color} 
\usepackage{natbib}

\newcommand{\degree}{^{\circ}} 
\newcommand{\eg}{\textit{e.g.}}

 \newcommand{\fig}[1]{Figure~\ref{fig:#1}}
 \newcommand{\figs}[2]{Figures~\ref{fig:#1} and \ref{fig:#2}}
 
 \newcommand{\sect}[1]{Section~\ref{sec:#1}}
 
 \newcommand{\sectss}[2]{Sections~\ref{sec:#1}--\ref{sec:#2}}

\journal{Advances in Space Research}

\begin{document}

%%%%%%%%%%%%%%%%%%%%%%%%%%%%%%%%%%%%%%%%%%%%%%%%%%%%%%%%%%%%%%%%%%%%%%%%%%%%%
%% Frontmatter
\begin{frontmatter}


% Use the tnoteref command within \title and fnref within \author or \address for footnotes;
% use the corref command within \author for corresponding author footnotes;
% \author{Name\corref{cor1}\fnref{label2}}
% \address{Address\fnref{label3}}

\title{Two successive partial mini-filament confined ejections}
%\tnotetext[footnote1]{This template can be used for all publications in Advances in Space Research.}

% Use optional labels to link authors explicitly to addresses:

\author[1]{M. Poisson\fnref{coni-fellow}}
\author[2]{C. Bustos}
\author[1]{M. L\'opez Fuentes\corref{cor}\fnref{coni-res}}
\cortext[cor]{Corresponding author}
\ead{lopezf@iafe.uba.ar}
\author[1,3]{C. H. Mandrini\fnref{coni-res}}
\author[1]{G.D. Cristiani\fnref{coni-res}}

\fntext[coni-fellow]{Fellow of CONICET, Argentina}
\fntext[coni-res]{Member of the Carrera del Investigador Cient\'\i fico, 
        CONICET, Argentina}

\address[1]{Instituto de Astronom\'{\i}a y F\'{\i}sica del Espacio, CC. 67
Suc. 28, CONICET-UBA, 1428 Buenos Aires, Argentina.}

\address[2]{Facultad de Ciencias Astron\'omicas y Geof\'\i sicas la Plata, Paseo del Bosque s/n, 
B1900 La Plata, Buenos Aires, Argentina.} 
 
\address[3]{Facultad de Ciencias Exactas y Naturales (FCEN), University of Buenos Aires (UBA), C1428EGA Buenos Aires, Argentina}

% Url can be given like this:
% \ead[url]{http://www.elsevier.com/wps/find/authorsview.authors/latex}

\begin{abstract}
Active region (AR) NOAA 11476 produced a series of confined plasma ejections, mostly accompanied by flares of X-ray class M, from 08 to 10 May 2012. The structure and evolution of the confined ejections resemble that of EUV surges; however, their origin is associated to the destabilization and eruption of a mini-filament, which lay along the photospheric inversion line (PIL) of a large rotating bipole. Our analysis indicate that the bipole rotation and flux cancellation along the PIL have a main role in destabilizing the structure and triggering the ejections. The observed bipole emerged within the main following AR polarity. Previous studies have analyzed and discussed in detail two events of this series in which the mini-filament erupted as a whole, one at 12:23 UT on 09 May and the other at 04:18 UT on 10 May. In this article we present the observations of the confined eruption and M4.1 flare on 09 May 2012 at 21:01 UT (SOL2012-05-09T21:01:00) and the previous activity in which the mini-filament was involved. For the analysis we use data in multiple wavelengths (UV, EUV, X-rays, and magnetograms) from space instruments. In this particular case, the mini-filament is seen to erupt in two different sections. The northern section erupted accompanied by a C1.6 flare and the southern section did it in association with the M4.1 flare. The global structure and direction of both confined ejections and the location of a far flare kernel, to where the plasma is seen to flow, suggest that both ejections and flares follow a similar underlying mechanism.  
\end{abstract}

\begin{keyword}
Solar Physics \sep  Solar activity \sep Solar flares \sep Solar ultraviolet emission \sep Solar X-ray and gamma-ray emission

%Los PACS vienen de https://publishing.aip.org/publishing/pacs/pacs-alphabetical-index#S
\PACS 96.60.-j \sep 96.60.Q- \sep 96.50.qe \sep 96.60.Tf, *96.60.tj \sep 96.60.Tf, *96.60.tk
\end{keyword}

\end{frontmatter}

\parindent=0.5 cm

%-------------Introduction------------------------------------------

\section{Introduction}
\label{sec:intro}

%\quad{\S\bf~Activity, ejecta, CMEs, jets, surges and sprays }\\
Solar activity produces several kinds of ejecta from full-fledge coronal mass ejections (CMEs) to surges, jets, and sprays. These ejections are characterized by their evolution, observed size, geometry, associated energies, and masses involved, and sometimes are classified considering the instruments with which they have been observed \citep{Webb12,Raouafi16,Canfield96,Schmieder96}. For instance, the classification of phenomena like surges and sprays has its origin in H$\alpha$ observations \citep[see e.g.][]{Foukal04}. The main difference between them is that while material in surges falls back down through the same magnetic structure through which it was ejected, in sprays the ejected mass apparently reaches escape velocity continuously ascending away from its point of origin in the chromosphere or very low corona \citep[see][]{Webb81}.

%\quad{\S\bf~Jets observations including mini-filaments}\\
Recent investigations have associated coronal jets to the eruption of so-called mini-filaments, whose plasma provides most of the ejected material \citep{Shen12,Adams14,Sterling15,Sterling16,Panesar16b,Joshi18,Yang18,Moore18}. Even in cases of lower resolution observations, mini-filament reformation and eruption has been postulated as the origin of a series of blow-out jets and related events \citep{Chandra17}. In all the previous works, the mechanism proposed for the destabilization of the mini-filament is the magnetic flux cancellation along the polarity inversion line (PIL) above which the mini-filament lays. Given the appropriate magnetic configuration, two reconnection processes may occur: a first, "inner", reconnection process in the magnetic field below the mini-filament, usually associated to a flare, and a second, "external" process produced by the ascent of the mini-filament, whose supporting magnetic field lines reconnect with the ones of the overlying magnetic field yielding the injection of material into open lines producing the jet \citep{Sterling16}.  

%\quad{\S\bf~Simulations with mini-filaments}\\
The role of mini-filaments in the process just described led \citet{Wyper17,Wyper18} to suggest that jets are produced by the same kind of mechanism, namely breakout reconnection, proposed to explain larger scale phenomena like CMEs \citep[see \eg ][]{Karpen12}. The numerical configuration in those simulations consists of a magnetic bipole emerging in a uniform unipolar magnetic flux area. This kind of configuration leads to the appearence of a coronal null-point above the bipole. Shearing motions imposed on the system provide free magnetic energy and produce a small flux rope that supports the mini-filament. The sheared magnetic structure expands and a reconnection process is triggered in the region around the coronal null point. As in the classical breakout model of CMEs, the reconnection removes part of the field restraining the flux rope, so it is propeled to rapidly ascend. At the same time, reconnection below the twisted flux rope is stimulated, consistently with the phenomenology described in the previous paragraph.  

%\quad{\S\bf~Filament partial ejections}\\
It has been observed in occasions that only part of the full structure of the filament is ejected \citep[see \eg ][and references therein]{Cheng18}. In those cases, the filament splits in two or more segments, some of which erupt at different times or not at all \citep{Zuccarello09,Tripathi06,Chandra10}. This behaviour has been identified also in the case of mini-filament eruptions associated to coronal jets \citep[see \eg ][]{Panesar17}.   

%\quad{\S\bf~The events studied in this paper }\\
The ejections studied in this article occurred in a similar magnetic configuration to that proposed to explain coronal jets, i.e. the presence of bipolar structures in the middle of a more extended unipolar region, in this case, the extended following polarity of active region (AR) NOAA 11576. A large magnetic bipole is observed to rotate along serveral tens of hours accompanied by a series of eruptions associated to flares. One of these, which occurred on 09 May 2012 at 12:23 UT, has been extensively analyzed by us in previous articles \citep{LopezFuentes15,LopezFuentes18}, and another, which occurred on 10 May 2012 04:18 UT, was studied by \citet{Yang18}. The main difference between the events studied here and a regular jet is that the ejected material remained confined within the closed magnetic structure of the AR instead of reaching open field lines. After the eruption, the ejected material ascended along field lines that connected the main polarities of the AR, part of it reached the farther footpoints of the lines, and part of it fell back to the region where the ejection initiated. This observed evolution resembled that of H$\alpha$ and EUV surges (see the aforementioned references). 

%\quad{\S\bf~The presence of a mini-filament}\\
The main EUV event analyzed here occurred in AR 11576 on 09 May 2012 and was associated to an M4.1 flare with a peak in GOES soft X-ray light curve at $\approx$ 21:01 UT. A close inspection of high resolution observations previous to the event, in the 304 \AA~band of the {\it Atmospheric Imaging Assembly} \citep[AIA:][]{Lemen12}, onboard the {\it Solar Dynamics Observatory} (SDO), indicates the presence of a mini-filament whose following eruption, accompanied by reconnection with the overlaying field, constitutes the main source of the ejected material. The extended analysis some tens of minutes before the main M4.1 event indicates that a previous smaller ejection ocurred 36 min before, involving the eruption of a different section of the same mini-filament. This previous event was associated to a less energetic C1.6 flare with peak intensity at $\approx$ 20:25 UT. 

%\quad{\S\bf~Roadmap}\\
In \sectss{obs_B}{euv_obs_stereo}, we analyze the evolution of both confined eruptions and accompanying flares in several wavelengths using data from the instruments described in \sect{data}. In \sect{comparison} we compare these successive events with the one analyzed by \citet{LopezFuentes18} and interperet our observations in the frame of that study. In \sect{ending} we present our concluding remarks.

%-----------DATA DESCRIPTION AND OBERVATIONS-------------------

\section{Observations and Characteristics of the Events}
\label{sec:obs}

%--------------------------------------------------------------

\subsection{The Data Used}
\label{sec:data}

\begin{figure}[]
\begin{center}
\includegraphics[width=0.9\textwidth]{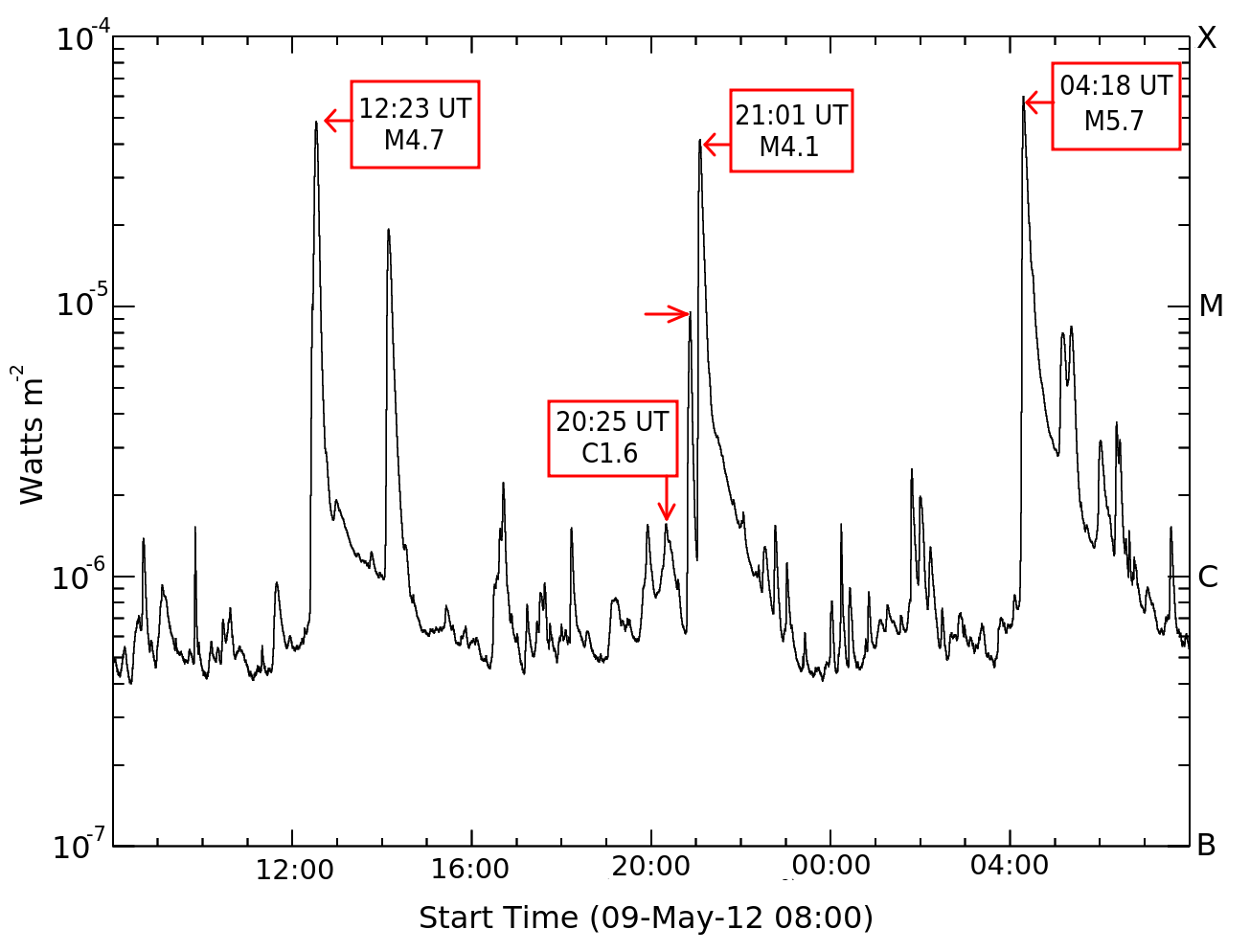}
\caption{GOES soft X-ray curve in the 1\,--\,8 \AA~channel showing the two flaring events after $\approx$ 20:20 UT on 9 May 2012. The arrows point to their peak times. Each arrow is accompanied by labels indicating the X-ray class and time of the flares. As a reference we identify the peak of the event at 12:23 UT that was studied by \citet{LopezFuentes18} and that of the event at 4:18 UT studied by \citet{Yang18}. The peak between the C1.6 and M4.1 flares, indicated with a red arrow, corresponds to a flare associated to the neighboring jet that occurred next to AR 11576 to the east at 20:45 UT and that is mentioned in \sect{obs_euv}.}
\label{fig:goes}
\end{center}
\end{figure}

%\quad{\S\bf~Instruments used and references}\\
To study the flares and confined eruptions we use UV continuum and EUV data from the {\it Atmospheric Imaging Assembly} \citep[AIA:][]{Lemen12}, onboard the {\it Solar Dynamics Observatory} (SDO), and EUV observations from the {\it Sun-Earth Connection Coronal and Heliospheric Investigation}  \citep[SECCHI:][]{Howard08}, onboard the {\it Solar Terrestrial Relations Observatory} (STEREO) spacecraft B, soft X-ray data from the {\it X-ray Telescope} \citep[XRT:][]{Golub07} onboard {\it Hinode}, and magnetograms from the {\it Helioseismic and Magnetic Imager} \citep[HMI:][]{Scherrer12}, onboard SDO. \fig{goes} depicts the soft X-ray curve in the 1\,--\,8 \AA~channel, where we have identified with arrows and labels both flaring events, as well as the one that occurred at 12:23 UT \citep{LopezFuentes18}, and the one that occurred at 04:18 UT on May 10 \citep{Yang18}. 

%\quad{\S\bf~Data description}\\
SDO/AIA data are from the 1700 \AA~channel (from now on AIA 1700, $T$ $\approx$ 5000 K) and 304 \AA~channel (AIA 304, $T$ $\approx$ 5$\times$10$^4$ K). Both events are observed in AIA 304, while only the flare at 21:01 UT is observed in AIA 1700. We select subimages containing AR 11576 from full-disk data for the temporal range covering the analyzed events. We coalign the subimages and construct movies that are associated to the figures of \sect{obs_euv}. All images are displayed in logarithmic intensity scales for better contrast. We complement the SDO/AIA EUV data with images of the events from the 195 \AA~channel of the SECCHI instrument. We use this band because it has the highest temporal resolution (5 minutes) in SECCHI. On the day of our events, STEREO-B was at an Earth ecliptic (HEE) longitude of -118$\degree$ away from Earth, from this location AR 11576 was seen in the solar limb. SDO/HMI data are line-of-sight (LOS) magnetograms, which are selected from full-disk images to study the evolution of the AR magnetic field (see \sect{obs_B}).

%----------------------------------------------------------------

\subsection{Summary of the Photospheric Magnetic Field Evolution}
\label{sec:obs_B} 

%\quad{\S\bf~General}\\
The magnetic evolution of AR 11476 from appearance, on the eastern solar limb, on 04 May 2012 until the end of 10 May has been discussed in detail by \citet{LopezFuentes18}. We summarize here their main results. 

\begin{figure}[]
\begin{center}
\includegraphics[width=0.8\textwidth]{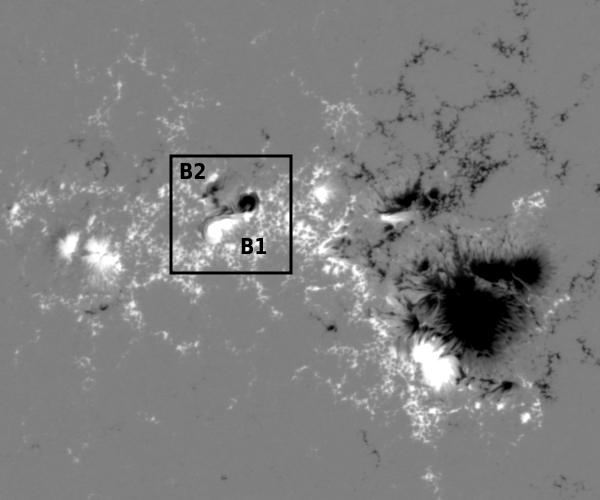}
\caption{SDO/HMI magnetogram of AR 11576 on 9 May at 20:00 UT, previous to the first event analyzed here. The field of view of the image covers approximately 260 Mm by 220 Mm. The location of bipoles B1 and B2 are indicated surrounded by a black box and corresponding labels. The events studied here occurred at the location of these bipoles.
}
\label{fig:hmi}
\end{center}
\end{figure}

\begin{figure}[]
\begin{center}
\includegraphics[width=1.0\textwidth]{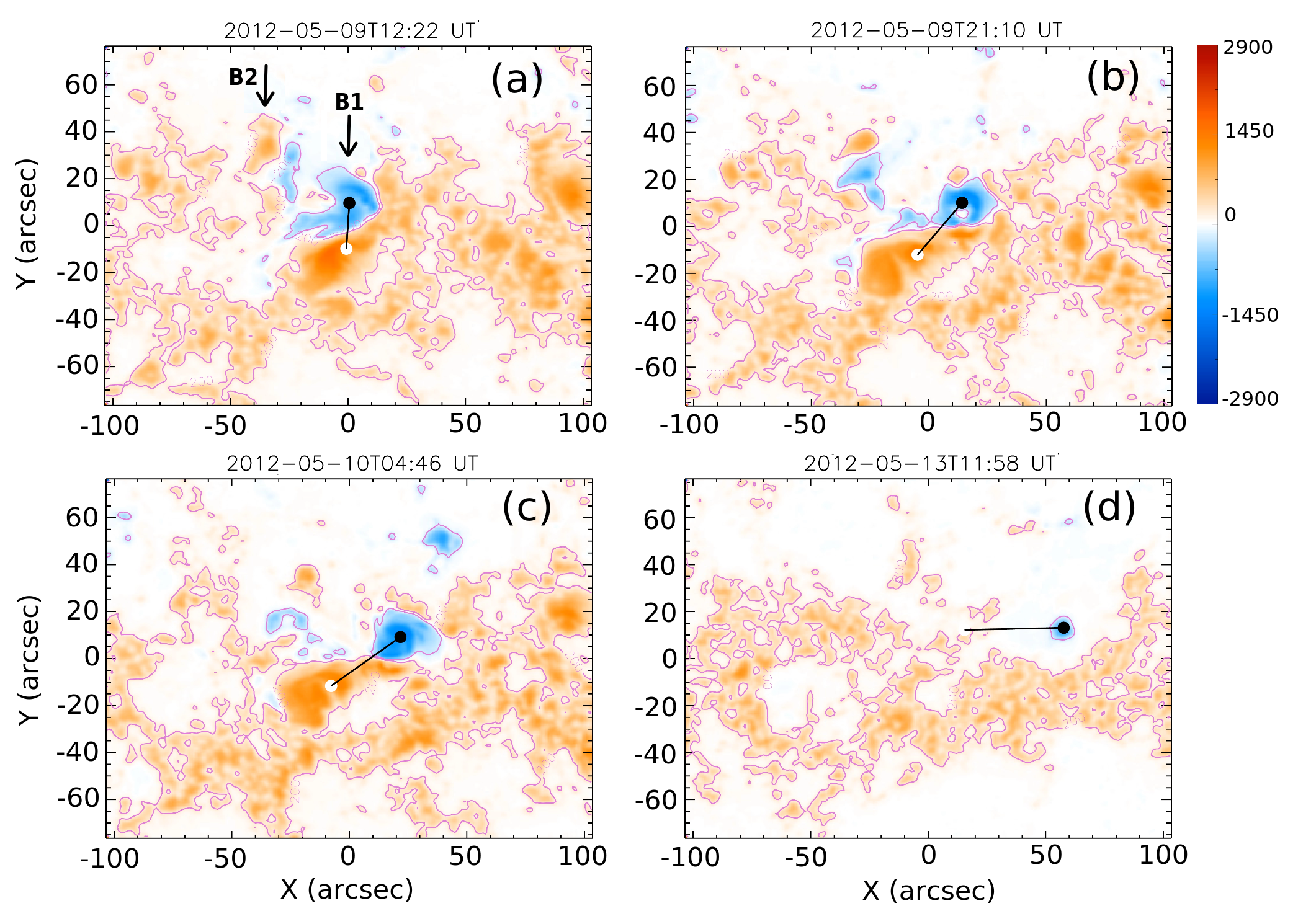}
\caption{SDO/HMI close-up magnetogram views of the bipoles shown in \fig{hmi}. The arrows in panel a point to the locations of the bipoles. Panels a\,--\,c correspond approximately to the times of the M flares on 9 and 10 May, while panel d shows the AR when B1 had almost completely disappeared. Blue corresponds to negative magnetic field and red to positive field. The magnetic field contours in pink correspond to $\pm$200 G. The axes are in arcsec. The black and white dots correspond to the location of the negative and positive magnetic barycenters of bipole B1, computed following the method described by \citet{Poisson15}. The evolution of the black segment joining the barycenters of B1 indicates the variation of the bipole tilt angle, defined here in the usual way, accordingly to Joy's law, as the bipole inclination with respect to the solar equator. A movie showing this rotation along three days is appended to this article as supplementary material (\href{run:./bipole-rotation.mov}{bipole-rotation.mov}).}
\label{fig:hmi_evol}
\end{center}
\end{figure}

%\quad{\S\bf~HMI evolution}\\
AR 11476 had a global bipolar structure; its preceding negative polarity was compact and was followed by a very dispersed following positive polarity (see \fig{hmi}). By 7 May two bipoles, which we have labeled as B1 and B2 in \fig{hmi}, were seen in the middle of the AR main positive polarity. The magnetogram in this figure corresponds to 20:00 UT, 25 min before the first event analyzed here. In the panels of \fig{hmi_evol} we show the bipole orientations at the times of particular events. \fig{hmi_evol}a corresponds roughly to the time of GOES soft X-ray peak of the M4.7 flare on 9 May at 12:23 UT, \fig{hmi_evol}b to approximately the time of GOES peak of the M4.1 flare analyzed in this article, \fig{hmi_evol}c depicts the bipole locations at approximately the time of an M5.7 flare peak at 04:18 UT on 10 May, while the last panel shows the magnetic configuration of the AR much later, when both bipoles had long ago stopped rotating and the negative polarity of B1 had almost disappeared. Both flares studied here and their related confined eruptions or surges, originated in the neighborhood of B1 and B2.

\begin{figure}[]
\begin{center}
\includegraphics[width=0.8\textwidth]{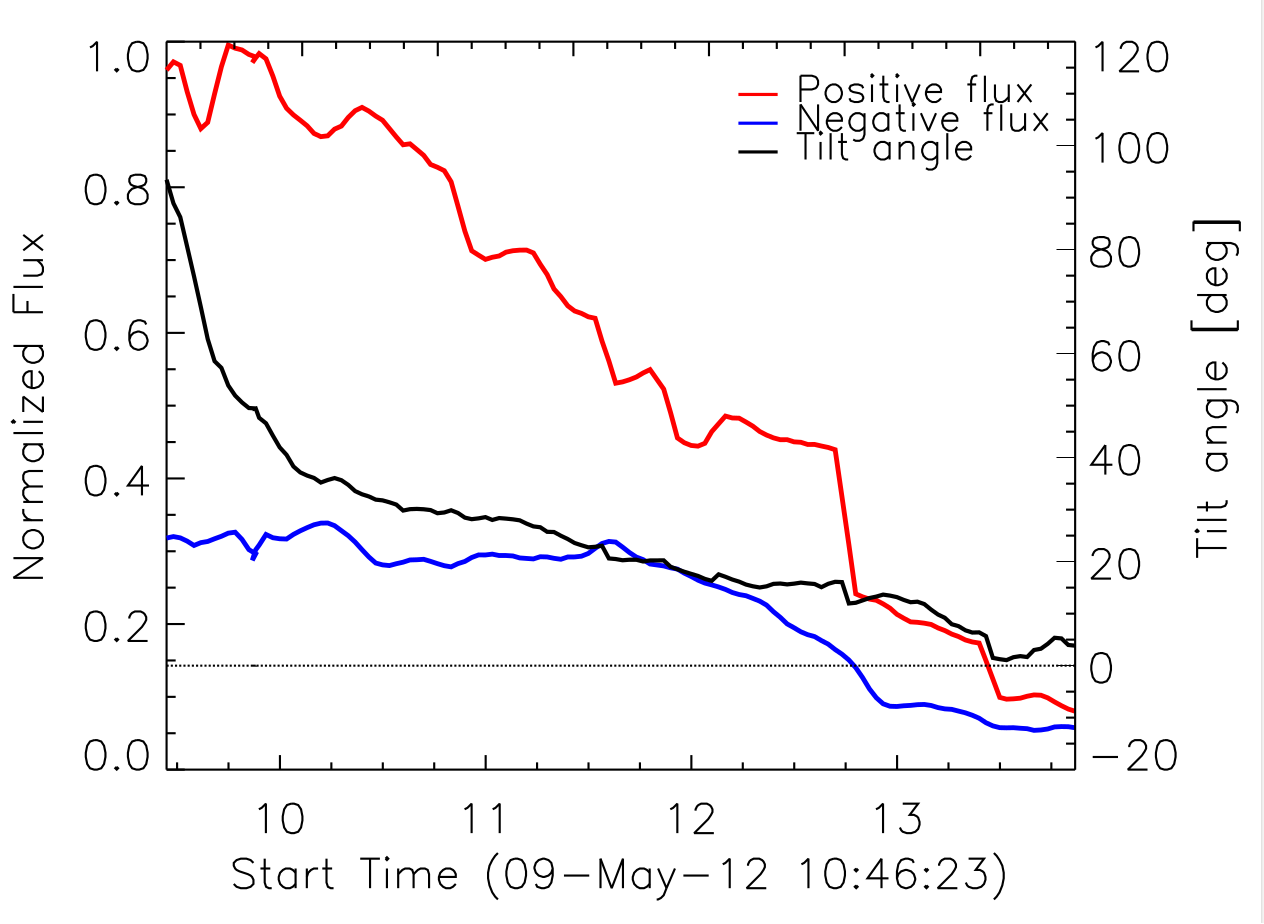}
\caption{Tilt angle rotation (black curve) and evolution of the magnetic flux of B1 (positive flux in red and negative in blue). Computations are done for values of the field above (below) 500 G (-500 G). 
}
\label{fig:rot_flux}
\end{center}
\end{figure}

%\quad{\S\bf~Bipole rotation and flux evolution}\\
The bipoles, B1 and B2 were seen rotating clockwise. B1 rotated by $\approx$ 180$\degree$ from 8 May at 00:00 UT to 10 May at 22:00 UT, while B2 rotated by $\approx$ 100$\degree$ from the same starting time until $\approx$ 10:00 UT on 10 May. The rotation of each bipole is characterized by the rotation of their tilt angle, which is determined following the method developed by \citet{Poisson15}. We include a movie showing the bipole rotations (\href{run:./bipole-rotation.mov}{bipole-rotation.mov}) as supplementary material. The bipole rotations imply the injection of magnetic helicity and consequently the accumulation of free magnetic energy; this evolution might contribute to the destabilization and eruption of the mini-filament during all the observed flares as proposed by \citet{Wyper17,Wyper18}. Furthermore, as we measured the bipole rotations, we also computed their magnetic flux. We found that the fluxes of B1 and B2 decreased steeply. In the case of B1 its flux mainly cancelled with the surrounding magnetic field, while in the case of B2 it was probably mostly dispersed. As discussed in \sect{intro}, magnetic flux cancellation has been observed associated with surges and jets. \fig{rot_flux} depicts the rotation of the tilt angle and the magnetic flux evolution of B1 from 8 May to 10 May. B1 is the largest and main bipole involved in the events studied in this article.

%-------------------------------------------------------------

\subsection{UV Continuum and EUV Evolution as Seen from Earth}
\label{sec:obs_euv}

\begin{figure}[]
\begin{center}
\includegraphics[width=0.8\textwidth]{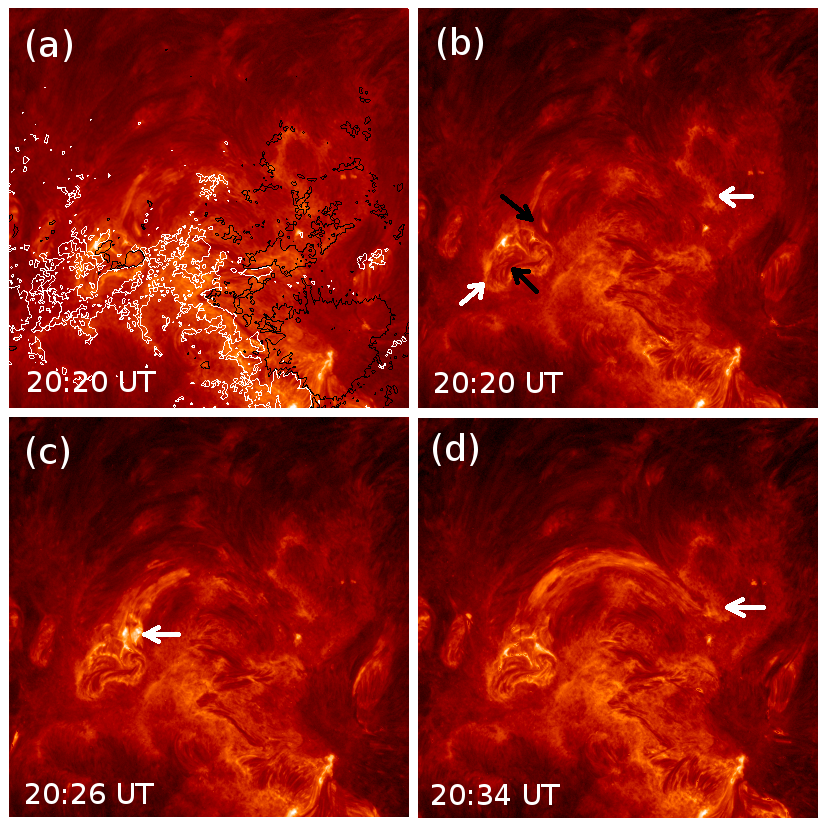}
\caption{SDO/AIA 304 images corresponding to the first partial eruption of the northern section of the mini-filament. Times are indicated in the panels. The panels are squares with a side length of 270 Mm. (a) Includes $\pm$ 200 G HMI contours as reference. Red (blue) color corresponds to positive (negative) magnetic field values. In (b) the mini-filament extending along the PIL of B1 is visible (see also \fig{aia304_zoom}b). Two black arrows point to the northern portion of the mini-filament, which is oriented in the NE--SW direction, and to its southern section, which is oriented in the NW--SE direction. The two white arrows indicate the location of the circular brightening and another one located on the negative AR main polarity to the W (see text for details). (c) Shows two bright kernels that we consider to be the footpoints of a short arcade or set of loops formed as the northen section of the mini-filament erupts. (d) Corresponds to the time of maximum extension of the surge associated to this partial mini-filament eruption. Notice that the loops along which the plasma flows have footpoints at the location of the elongated brightening to the W. A movie showing the AIA 304 evolution of the C1.6 and M4.1 flares, as well as the partial mini-filament eruptions, is attached as electronic supplementary material (\href{run:./aia304.mp4}{aia304.mp4}). 
}
\label{fig:aia304a}
\end{center}
\end{figure}

%\quad{\S\bf~AIA 304 general description}\\
\figs{aia304a}{aia304b} show AIA 304 images at the times of the first partial mini-filament eruption and accompanying C1.6 flare, and of the M4.1 flare and second partial mini-filament eruption, respectively. An HMI magnetogram contour is overlaid on the images in \fig{aia304a}a and \fig{aia304b}a as a reference. 

As discussed in \sect{intro}, recent observations relate jets to the eruption of mini-filaments, which are smaller versions of normal large-scale filaments. In the events analyzed in this article we find evidence of the partial eruption of a curved mini-filament that extends all along the PIL of B1. This mini-filament is observed to erupt and reform several times from 8 May to 10 May. We infer that the continuous rotation and flux cancellation of B1 along these days contribute to the mini-filament reformation and destabilization. We include as supplementary material a movie with high temporal resolution (5 images {\it per} minute) in AIA 304 showing both flares and mini-filament eruptions (\href{run:./aia304.mp4}{aia304.mp4}). In the movie a jet is also observed at the east of the AR (right at the left border of the images) occurring around 20:45 UT. This event is unrelated to those studied here as it occurred away from the bipole areas where the mini-filament eruptions and flares took place. A flare associated to this jet is identified in \fig{goes} with a red arrow (no label) between the C1.6 and M4.1 flares.

\begin{figure}[]
\begin{center}
\includegraphics[width=0.6\textwidth]{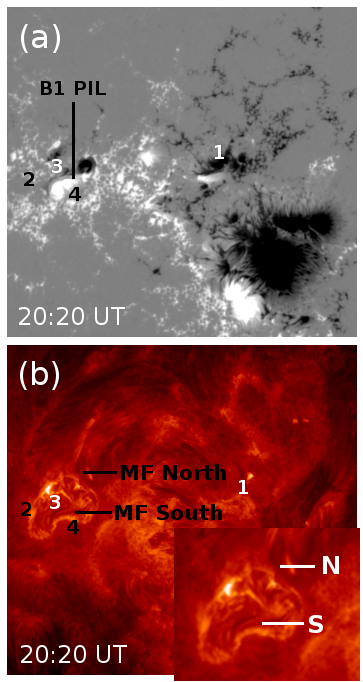}
\caption{Zoom of a portion of \fig{aia304a}a and b. (a) 20:20 UT HMI magnetogram shown as reference. We number the main polarities associated to the development of the analyzed flares and surges as described in \sect{comparison}. The polarity inversion line (PIL) of bipole B1 is indicated. (b) Location of the same polarities on the cotemporal AIA 304 image. We point the locations of the northern and southern portions of the erupting mini-filament (indicated as MF). We include a further enlargement of the big bipole area to easily identify both mini-filament portions, indicated as N (north) and S (south). 
}
\label{fig:aia304_zoom}
\end{center}
\end{figure}

\begin{figure}[]
\begin{center}
\includegraphics[width=0.7\textwidth]{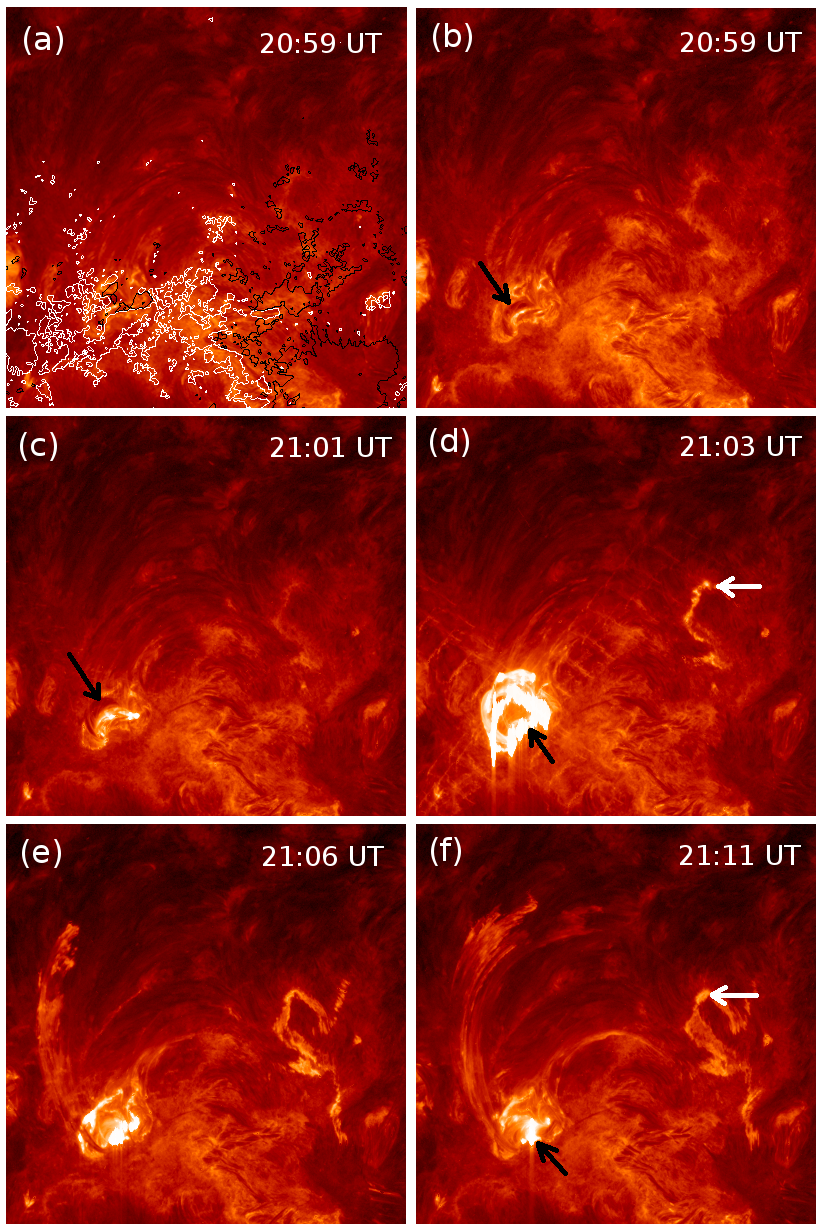}
\caption{Second surge and associated M4.1 flare evolution as observed in AIA 304. (a) Includes $\pm$ 200 G HMI contours as reference. Red (blue) color corresponds to positive (negative) magnetic field values. The panels are squares with a side length of 270 Mm. The black arrows in (b) and (c) indicate the location of the erupting mini-filament. The white arrows in (d) and (f) indicate the location of a far flare kernel on the negative main polarity of the AR. Notice the ascending material in (e) and (f) that follows the AR loops connecting with the far kernel. The observed evolution can be more clearly followed in the accompanying movie (\href{run:./aia304.mp4}{aia304.mp4}).}
\label{fig:aia304b}
\end{center}
\end{figure}

%\quad{\S\bf~Figures description - 304 first eruption}\\
\fig{aia304a} illustrates the initiation and evolution of the first event. High resolution imagers, like SDO/AIA, allow us to clearly observe a curved dark structure with one section oriented in the NE--SW direction (henceforth, the northern section) and another oriented in the NW-SE direction (henceforth, the southern section, see also \fig{aia304_zoom}b). These two sections have been pointed with black arrows in \fig{aia304a}b. \fig{aia304a}c shows the appearance of two bright kernels (indicated with a white arrow) that we consider to be the footpoints of a short arcade formed as the mini-filament northern section erupts. At the same time, a dense plasma eruption is seen extending towards the W. \fig{aia304a}d shows the eruption at its maximum extension reaching a far kernel, indicated with a white arrow, that globally has the same shape as the far kernel observed during the earlier event at 12:23 UT. Another elongated brightening, pointed by a white arrow in Figure 5b and d, is also seen to the W on the main negative AR polarity. The location of the first flare kernels, as well as the place from where the mini-filament eruption initiates, indicates that only the northern section of the mini-filament erupts.

%\quad{\S\bf~Figure 6 and the circular shape brightening}\\
In \fig{aia304_zoom} we show an enlargement of the image in \fig{aia304a}a and b. \fig{aia304_zoom}a shows the magnetic map as a reference. In this panel we have numbered the main polarities involved in the events studied in this article. \fig{aia304_zoom}b corresponds to the AIA 304 image where the shape of the mini-filament before both eruptions can be more clearly seen. For an easier identification of the mini-filament portions we added an inset showing a further enlargement of the bipole area. The north and south portions for the mini-filament are indicated both in \fig{aia304_zoom}b and in the inset. Notice the presence of a circular-shape brightening, pointed by a white arrow in \fig{aia304a}b (also observable in \fig{aia304_zoom}b), that surrounds all the negative and part of the postitive polarity of B1 and all of B2, as happened in the event at 12:23 UT. This circular brightening is present since around 08 May, which strongly suggests that energy release at a very low rate occurred accompanying the rotation of the bipoles plausibly because of their interaction with the overlaying magnetic field.

\begin{figure}[]
\begin{center}
\includegraphics[width=0.8\textwidth]{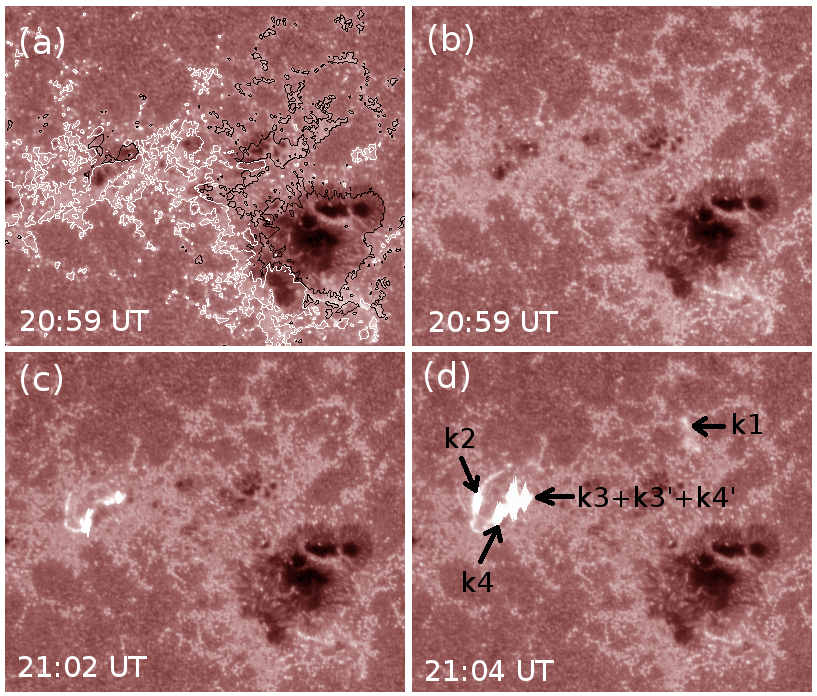}
\caption{SDO/AIA 1700 images of the M4.1 flare. Times are indicated in the panels. (a) Includes $\pm$ 200 G HMI contours as reference. Red (blue) color corresponds to positive (negative) magnetic field values. The panels are squares with a side length of 270 Mm. (b)--(f) Show the evolution of the flare kernels which can be identified in this band. (d) Shows the flare kernels numbered using the same letters as in \citet{LopezFuentes18}. A movie showing the AIA 1700 evolution of the flare is attached as electronic supplementary material (\href{run:./aia1700.mp4}{aia1700.mp4}).
}
\label{fig:aia1700}
\end{center}
\end{figure}

%\quad{\S\bf~Figures description - 304 second eruption}\\
\fig{aia304b} shows the evolution of the second surge and associated M4.1 flare as observed in the AIA 304 band. As in \fig{aia304a}, in \fig{aia304b}a we include an SDO/HMI contour as a reference to easily identify the location of the main AR polarities and B1 and B2 bipoles. In \fig{aia304b}b we indicate with a black arrow the remaining southern portion of the mini-filament, which is seen prominently right before starting to rise. In \fig{aia304b}c, the mini-filament is observed already ascending and an incipient brightening is noticeable below it. In \fig{aia304b}d, all the area around B1 is seen bright as the peak intensity of the flare happens. The mini-filament material is ejected and it is part of the ascending bright front whose evolution is seen in the following panels (e and f). The full evolution can be followed in the AIA 304 movie accompanying this article as supplementary material (\href{run:./aia304.mp4}{aia304.mp4}). Notice also the far brightening located at the AR main negative polarity and indicated with white arrows in \fig{aia304b}d and f. As this area brightens consequently with the M4.1 flare development, we infer that it is another kernel of the flare, as we show in the description of \fig{aia1700} in what follows. As observed in \fig{aia304b}e and f, and in the accompanying movie, the ejected material travels along magnetic loops connecting the area surrounding the bipoles with this far kernel. 

%\quad{\S\bf~Velocity estimation from the 304 images}\\
From the observed evolution of the ascending surge front in AIA 304 data we can estimate its velocity in the plane of the sky. By following the vertical motion of the bright front from the original location of the mini-filament to the maximum volumetric extent of the surge we estimate a mean velocity of approximately 280 km s$^{-1}$. We obtained this velocity by tracking the front along successive images and then computing the slope of the position versus time curve, from the original filament location to the farthest position reached by the surge front. Although this velocity is higher than the 190 km s$^{-1}$ obtained for the 12:23 UT surge studied by \citet{LopezFuentes18}, it is still within the usual values observed in this kind of events \citep{Raouafi16}.

%\quad{\S\bf~Figures description - UV Continuum}\\
In order to identify the locations of the main M4.1 flare kernels, we analyze AIA 1700 images. The C1.6 flare is not visible in this AIA band. \fig{aia1700} shows a series of images in AIA 1700. In \fig{aia1700}b--d the flare kernels and their evolution is clearly seen. In \fig{aia1700}d we have labeled the different kernels using the same letters as \citet{LopezFuentes18} used in their analysis of the event at 12:23 UT. The location of the initial emission in this band, as well as its evolution, together with the AIA 304 evolution discussed in the previous paragraph, suggests that it is the southern section of the mini-filament that erupts in association with the M4.1 flare. Notice that the far kernel to where the plasma is observed to flow in AIA 304 during this second mini-filament eruption is barely seen in this band. We include as supplementary material a movie with the highest temporal resolution in AIA 1700 showing the M4.1 flare and the second mini-filament partial eruption (\href{run:./aia1700.mp4}{aia1700.mp4}); in this movie we can see that some plasma appears flowing up at this temperature range. Notice the bright kernels associated to the jet at the east of the AR (right at the left border of the images) that starts at approximately 20:45 UT. As we mentioned before, this event is unrelated to those studied here. 

\begin{figure}[]
\begin{center}
\includegraphics[width=0.7\textwidth]{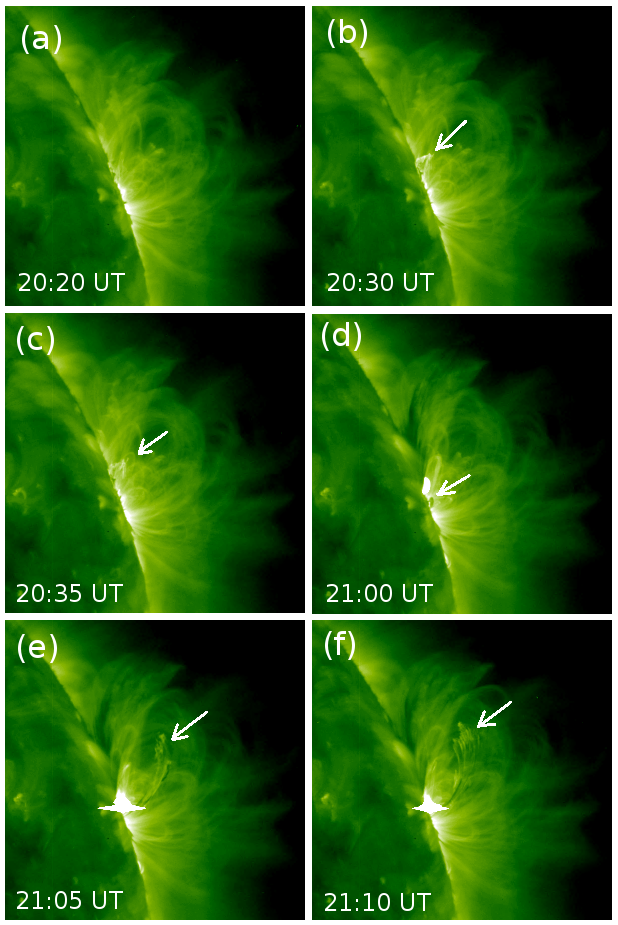}
\caption{STEREO B observations obtained with the SECCHI instrument in the 195 $\AA$ channel on 9 May 2012. (a) Corresponds to the location and illustrates the coronal loop structure of AR 11576, as seen on the limb from the STEREO B point of view before the events studied here. The cadence of these observations is 5 min. The panels are squares with a side of $\approx$340 Mm. The times of the images are indicated in the panels. In (b) and (c) we indicate with white arrows the location of the first surge front as it erupts. In (d) we use a white arrow to identify the location of the mini-filament portion associated to the second surge. In (e) and (f) the front of the second surge is indicated with white arrows 5 min and 10 min after the eruption begins. The associated M4.1 flare is also observable as an extended area of saturated pixels at the base of the ejected structure. The full evolution observed from STEREO B can be seen in the accompanying movie (\href{run:./stereo195.mp4}{stereo195.mp4}).}
\label{fig:stereo}
\end{center}
\end{figure}

%\quad{\S\bf~Other EUV surges in the AR}\\
As mentioned in \citet{LopezFuentes18}, several EUV surges can be identified in AIA 304 when examining images in this band in Helioviewer (www.helioviewer.org). All these surges are associated to M-class flares, except the first partial eruption studied in this article at $\approx$ 20:25 UT. The more extended surges are: one on 8 May at $\approx$13:05 UT (flare M1.4), another one on 9 May at $\approx$ 12:23 UT \citep[flare M4.7, see][]{LopezFuentes18}, the one analyzed here at $\approx$ 21:01 UT (flare M4.1), and two on 10 May at $\approx$ 04:15 UT \citep[flare M5.7, see][]{Yang18} and at $\approx$ 20:25 UT (flare M1.7). All these M-class flares and surges started mainly at the location of the largest rotating bipole (B1) and the plasma was launched along large-scale loops with far footpoints on the negative main AR polarity to the W. This series of events suggests a recurrence of energy storage, mini-filament reformation as proposed by \citet{Chandra17} \citep[see also the simulations of][]{Wyper17,Wyper18}, and energy release processes in a time range between around 8 and 23 hours.

%---------------------------------------

\subsection{EUV Evolution as Seen by STEREO-B}
\label{sec:euv_obs_stereo}

%\quad{\S\bf~General description}\\
As described in \sect{data}, at the time of the studied events the STEREO probe B was located in an orbital position from which AR 11576 was observable on the solar limb. This provides us the vantage point of a side view to observe the ejections. In \fig{stereo} we show a series of SECCHI images in the 195\AA~band (hereafter SECCHI 195) covering both surge evolutions. The cadence of this set of images is 5 min. In \fig{stereo}a we show an image of AR 11576 right before the first event took place (at 20:20 UT). By 20:30 UT (\fig{stereo}b) we identify a bright ascending front, indicated with a white arrow, that coresponds to our first surge. It can be still observed progressing in \fig{stereo}c (20:35 UT), where we identified the highest point of the surge front with a white arrow. Regarding the evolution of the second surge, in \fig{stereo}d (21:00 UT), we point with a white arrow the location of the mini-filament portion associated to this eruption, as observed from STEREO. In \fig{stereo}e and f (21:05 UT and 21:10 UT, respectively), the white arrows indicate the position of the ascending surge front. Notice the bright saturated pixels at the base of the erupting material produced by the associated M4.1 flare. The evolution can be fully observed in the short movie accompanying this article as supplementary material (\href{run:./stereo195.mp4}{stereo195.mp4}).

%\quad{\S\bf~Velocity measured from STEREO data}\\
Although the cadence of the SECCHI 195 data does not allow a continuous track of the erupted material as in the case of AIA 304, we can still observe the mini-filament just before the ejection begins and the ejected material at a high altitude 5 min later. Therefore, we can make a very rough estimation of the mean velocity of the surge by dividing the observed distance traveled by the material by the time difference between the images in \fig{stereo}d and f. The mean velocity estimated in this way is approximately 300 km s$^{-1}$, which is consistent with the result obtained from the AIA 304 analysis. 

%---------------------------------------

\begin{figure}[]
\begin{center}
\includegraphics[width=0.7\textwidth]{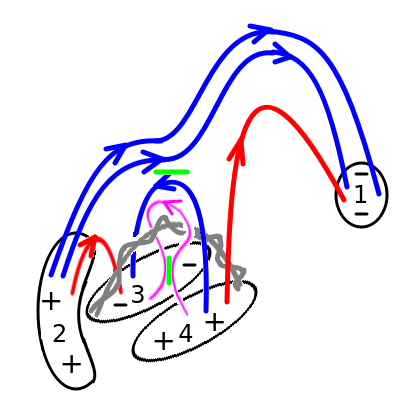}
\caption{3D sketch showing sets of field lines connecting the different polarities identified in \fig{aia304_zoom} and the relative location of the mini-filament. As the mini-filament rises two reconnection processes occur, as identified with thick green segments: the internal one below the mini-filament and the external one above it, which reconnects (blue) field lines connecting polarities 3 and 4 with field lines connecting 1 and 2. This latter process eventually yields the injection of mini-filament material in field lines connecting 3 and 2 and 4 and 1 (highlighted in red color) producing the observed eruption. See \sect{comparison} for a detailed description.
}
\label{fig:sketch}
\end{center}
\end{figure}

\section{Comparing the phenomenology of the confined ejections on 9 May 2012}
\label{sec:comparison}

%\quad{\S\bf~General description}\\ 
In this section we compare the confined eruptions of 9 May 2012, which were associated to the M4.7, the C1.6, and M4.1 flares occurring at 12:23, 20:25, and 21:01 UT, respectively, as identified in \fig{goes}. In \citet{LopezFuentes18} we performed a detailed study of the magnetic field connectivity in the regions where the flare kernels were located and interpreted our observational results in terms of the magnetic field topology of AR 11576. We interpret the observations of the two successive flares and confined eruptions, just discussed, in the context of our previous results in view that the magnetic field configuration has not changed substantially, i.e. the main change is the continuous rotation of B1 and B2. 

%\quad{\S\bf~First fainter flare and ejection}\\
It is clear from \fig{aia304a} that only the northern section of the mini-filament erupts at around 20:25 UT, when two small flare kernels are seen at both sides of the northern section of the PIL of B1. A C1.6 flare accompanies the eruption; this is a faint event compared to the other ones on 9 May, in particular, it is not even classified as a flare in SolarMonitor (https://www.solarmonitor.org/). Based on the magnetic field topology computed in \citet{LopezFuentes18}, the evolution seen in the movie that accompanies this article (\href{run:./aia304.mp4}{aia304.mp4}), and the different panels in \fig{aia304a}, we infer that the continuous rotation of B1 and flux cancellation at its PIL, produces an instability triggering this ejection. When this happens, the mini-filament starts to rise and two reconnection processes set in. Magnetic field lines connecting polarities 3 and 4 (see \fig{aia304_zoom}a), at both sides of the PIL, reconnect below the mini-filament, in what is called internal reconnection. \fig{sketch} shows a 3D sketch of the magnetic field lines connecting the different polarities and the relative position of the mini-filament. The internal reconnection site is identified with a thick green segment below the mini-filament location in \fig{sketch}. This process creates a new set of field lines whose footpoints are located at both sides of the PIL, clearly seen in \fig{aia304a}c as the two bright kernels pointed with a white arrow. The other set of reconnected field lines surrounds the mini-filament. Simultaneously, and as the mini-filament rises, the closed field lines above it that still connect 3 and 4 are forced to reconnect with those connecting 2 and 1. This reconnection process is called external and is identified with a thick green segment above the mini-filament location in \fig{sketch}. In this process the mini-filament plasma has access to the large-scale field lines connecting 4 and 1 and we observe the surge flowing towards the southern portion of the far kernel to the W in 1, pointed with a white arrow in \fig{aia304a}d. Notice that in this case the loops connecting 4 and 1 are shorter and probably not as high as those in the surge observed associated to the flare at 12:23 UT and the later one at 21:01 UT. Because of the surge and other brightenings, the counterpart kernel in 4 is not evident in AIA images. At the same time we should have seen part of the mini-filament plasma reaching a flare kernel located in 2, in a similar way as happens with the surge at 12:23 UT; however as these events are less energetic than the latter one, this is not visible in the 304 movie. Anyway, a flare kernel is located in 2 at around 20:23 UT and later (see the white arrow to the east in \fig{aia304a}c. Summarizing, this fainter flare and confined ejection can be explained in a similar way as the events at 12:23 UT, but in this particular case only the northern portion of the mini-filament is ejected. 

%\quad{\S\bf~Second flare and ejection}\\
As we show in \fig{aia304b} the origin of the events at 21:01 UT was the ejection of the southern portion of the mini-filament on the PIL of bipole B1. In this particular case, the flare and eruption are homologous to the events studied by \citet{LopezFuentes18}. In \fig{aia1700}b we have labeled the flare kernels as k1, k2, k3 and k3', and k4 and k4', using the same notation as in that article. We associate these kernels to the location of the polarity regions labeled as shown in \fig{aia304_zoom}a. Kernels k3'and k4' are located in polarities 3 and 4 and are the result of the internal reconnection process that sets in when the southern portion of the mini-filament destabilizes and starts rising. Kernels k4 and k1, as well as kernels k2 and k3, result from the external reconnection process when the loops above the rising mini-filament and linking polarities 3 and 4 reconnect with the overlying closed field lines connecting 2 and 1 (see \fig{sketch}). 

%\quad{\S\bf~Sketch valid for both events}\\
Finally, we notice that, even with the high spatial resolution of AIA images, it is not clear if the mini-filament at the time of the events studied here is formed by two sections and, therefore, we have no clue of why the mini-filament erupts in two sections.    

%---------------------------------------
\section{Concluding Remarks}
\label{sec:ending}

We studied a two-step mini-filament eruption in AR NOAA 11576 on 9 May 2012, at 20:25 UT and 21:01 UT. These eruptions were observed as EUV surges in SDO/AIA data and occurred in conjunction with a C1.6 and an M4.1 flares, identified in the GOES soft X-ray emission curve. We studied the evolution of the photospheric magnetic field of the AR, observed in SDO/HMI magnetograms, and related these events to the rotation and flux cancellation of bipolar structures located in the middle of the AR main positive polarity. 

By identifying the flare kernels and comparing them with the previous event at 12:23 UT, in \sect{comparison} we propose a phenomenological explanation of both eruptions. This is done in terms of the ejection of two sections of a  mini-filament located along the PIL of the largest rotating bipole, magnetic connections between the magnetic polarities of the AR configuration, and reconnection processes between field lines connecting these polarities. The location of flare kernels, the shape of the brightened structures, and the observed evolution suggest that the same magnetic topology found by \citet{LopezFuentes18} is present during the events studied in this article. 

Except for the fact that the mini-filament material is ejected through closed magnetic field lines that connect the main polarities of the AR, the studied events resemble the kind of evolution observed and modeled in previous works for jets \citep[see \eg ][ and other references in the Introduction]{Sterling16,Wyper17,Wyper18}, for which the material is ejected along open magnetic field lines. Our study confirms the role of mini-filaments in the development of jets and surges and contributes to understand the magnetic configurations and evolution associated to this kind of events. In particular, we further confirm the suggestion proposed in previous works (see references in \sect{intro}), that the repeated mini-filament reconstructions and eruptions accompanied by flares, as observed in AR 11576, are sustained by the continuous rotation and flux cancellation of the bipolar structures located in the middle of the AR main positive polarity. Although our study focuses on events of the surge type, in which the ejected material falls back to the coronal base, the similarity of the processes producing the events analyzed here and jet observations (both standard and blow-out) studied elsewhere, suggest the possibility that other typical ejections, such as H$\alpha$ sprays (see references in \sect{intro}), might be produced by similar mechanisms. This possibility would need to be addressed in future high resolution studies of solar ejective phenomena.

\section*{Acknowledgements}
The authors thank the anonymous reviewers for useful comments and suggestions. MLF, CHM and GC are members of the Carrera del Investigador Cient\'{\i}fico of the Consejo Nacional de Investigaciones Cient\'{\i}ficas y T\'ecnicas (CONICET) of Argentina. MP is a CONICET Fellow. MP, MLF, GC and CHM acknowledge financial support from the Argentinean grant PICT 2012-0973 (ANPCyT).

\bibliography{paper_surge2}
\bibliographystyle{elsart-harv}

\end{document}